\documentclass[twocolumn,showpacs,aps,prl,superscriptaddress]{revtex4}

\usepackage{graphicx}
\usepackage{dcolumn}
\usepackage{amsmath}
\usepackage{epsfig}

\input pubboard/babarsym
\def\BRbztodstdstNum {\ensuremath{(8.3 \pm 1.6(stat) \pm
1.2(syst))\times 10^{-4}}}
\def\BRbztodstdst {\ensuremath{{\BR}(\Bztodstdst) = \BRbztodstdstNum}}

\def\fsideVal {1.72 \xspace}
\def\fsideErr {0.10 \xspace}

\def\OnResLumi {20.4\invfb}

\def\NBB {(22.8 \pm 0.4) \times 10^{6}}

\def\DelELowSide {50\mev}
\def\DelEHiSide {200\mev}
\def\mesLowSide {5.20\gevcc}
\def\mesMidSide {5.26\gevcc}
\def\mesHiSide  {5.29\gevcc}

\def\Dstarp     {\ensuremath{D^{*+}}\xspace}
\def\Dstarm     {\ensuremath{D^{*-}}\xspace}

\def\Dp         {\ensuremath{D^+}\xspace}
\def\Dm         {\ensuremath{D^-}\xspace}

\def\BztoDDbar	{\ensuremath{B^0 \to D^{(*)+} D^{(*)-}}\xspace}

\def\DstDst {\ensuremath{D^{*+} D^{*-}\xspace}}
\def\Dzpip{\ensuremath{D^0 \pi^+ \xspace}}

\def\Dppiz{\ensuremath{D^+ \pi^0 \xspace}}
\def\Dmpiz{\ensuremath{D^- \pi^0 \xspace}}

\def\Dstptopip  {\ensuremath{\Dstarp \to \Dz\pip}\xspace}
\def\Dstptopiz  {\ensuremath{\Dstarp \to \Dp\piz}\xspace}

\def\DeltaEStd  {\ensuremath{\Delta E} \xspace}
\def\SsqovSpB   {\ensuremath{S^2/(S+B)}\xspace}

\def\chisqM     {\ensuremath{\chi^2_{Mass}}\xspace}

\newcommand{\BABARPubYear}    {01}
\newcommand{\BABARPubNumber}  {24}

\newcommand{\SLACPubNumber} {9152}

\def\figurebox#1#2#3{%
    \def\arg{#3}%
    \ifx\arg\empty
    {\hfill\vbox{\hsize#2\hrule\hbox to #2{\vrule\hfill\vbox to #1{\hsize#2\vfill}\vrule}\hrule}\hfill}%
    \else
    {\hfill\epsfbox{#3}\hfill}%
    \fi}

\begin{document}

\preprint{\babar-PUB-\BABARPubYear/\BABARPubNumber} 
\preprint{SLAC-PUB-\SLACPubNumber} 

\begin{flushleft}
\babar-PUB-\BABARPubYear/\BABARPubNumber\\
SLAC-PUB-\SLACPubNumber\\[5mm]
\end{flushleft}

\title{
{\large \bf
Measurement of the Branching Fraction and \CP\ Content for the Decay 
\Bztodstdst} 
}

%
\author{B.~Aubert}
\author{D.~Boutigny}
\author{J.-M.~Gaillard}
\author{A.~Hicheur}
\author{Y.~Karyotakis}
\author{J.~P.~Lees}
\author{P.~Robbe}
\author{V.~Tisserand}
\author{A.~Zghiche}
\affiliation{Laboratoire de Physique des Particules, F-74941 Annecy-le-Vieux, France }
\author{A.~Palano}
\author{A.~Pompili}
\affiliation{Universit\`a di Bari, Dipartimento di Fisica and INFN, I-70126 Bari, Italy }
% ...
\collaboration{The \babar\ Collaboration}
\noaffiliation

\date{\today}

\begin{abstract}
We report a measurement of the branching fraction of the decay
$B^0 \to D^{*+} D^{*-}$ and of the $C\!P$-odd component of its final 
state using the \mbox{\slshape B\kern-0.1em{\smaller A}\kern-0.1em 
B\kern-0.1em{\smaller A\kern-0.2em R}} detector. 
With data corresponding to an integrated luminosity 
of $20.4\,{\rm fb^{-1}}$ collected at the $\Upsilon(4S)$ resonance
during 1999-2000, we have reconstructed 38 candidate signal events in
the mode $B^0 \to D^{*+}D^{*-}$ with an estimated background of $6.2
\pm 0.5$ events.  From these events, we determine the branching fraction to be
$\BR(B^0 \to D^{*+}D^{*-}) = (8.3 \pm 1.6(stat) \pm 1.2(syst)) \times 10^{-4}$.
The measured $C\!P$-odd fraction of the final state is
$0.22 \pm 0.18(stat) \pm 0.03(syst)$.

\end{abstract}


\maketitle
After the observation of time-dependent \CP-violating asymmetries in
the decays of neutral $B$ mesons to \CP eigenstates containing
charmonium~\cite{ref:babars2b,ref:belles2b}, it is interesting to
extend the search
for \CP-violating effects to Cabibbo-suppressed double charm modes,
such as \BztoDDbar~\cite{ref:cc}.
The interference of the dominant tree amplitude with the mixing
diagram is sensitive to the angle $\beta$ of the Unitarity Triangle in this
case as well; however, the theoretically uncertain contribution of
penguin amplitudes with different weak phases is potentially
significant and may shift the observed asymmetry by an amount that
depends on the ratio of the penguin and tree contributions and their
relative weak phases.
The \Bztodstdst vector-vector final state has very clear experimental
signatures that make it an interesting candidate for \CP-violation
measurements.  However, it is not a pure \CP eigenstate and a dilution
of the measured asymmetry can be produced by a $P$-wave, \CP-odd,
component. A time-dependent angular analysis of the decay
products~\cite{ref:dunietz} can
remove the dilution by resolving the \CP-even and \CP-odd components.
As a precursor to
measuring time-dependent \CP-violating asymmetries using the decay
\Bztodstdst, we report in this letter a measurement of the \Bztodstdst
branching fraction and a measurement of the \CP-odd component,
$R_\perp$,
of the final state.  These measurements represent significant improvements
over the previous measurements $\BR(\Bztodstdst) =
(9.9^{+4.2}_{-3.3}(stat)\pm1.2(syst))\times 10^{-4}$ and $(1-R_\perp) <
0.11$ at 90\% C.L.~\cite{ref:cleoprd62}.

The data used in this analysis were collected with the \babar\
detector~\cite{ref:babar} at the \pep2\ storage ring~\cite{ref:pepii}
located at the Stanford Linear Accelerator Center.  
This data sample represents an integrated
luminosity of \OnResLumi collected on the \FourS
resonance.
Assuming 50\% of the \FourS decays give \BzBzb, the
number of neutral $B$ mesons in this sample is $\NBB$.

Charged particles are detected and their momenta measured with the
combination of a 40-layer drift chamber (DCH) and a five-layer silicon
vertex tracker (SVT), both operating in a 1.5\,T solenoidal magnetic field.
The charged particle tracking system
allows particles with low momentum in the laboratory frame 
to be reconstructed efficiently, a property that
is very important for this analysis.
This efficiency begins to turn on at a momentum of $\sim$60\mevc and
reaches its maximum value at around 200\mevc.
Photons are detected by a CsI(Tl) electromagnetic calorimeter (EMC) that
provides high detection for energies above 20\mev, with typical energy
and angular resolutions of 3\% and 4\,mrad, respectively, for 1\gev photons.
Charged particle
identification is provided by the ionization loss measurements
in the SVT and DCH, and by an internally reflecting ring-imaging
Cherenkov detector (DIRC) covering the central region of the detector.


Events are selected by requiring
three or more charged tracks and that the normalized second Fox-Wolfram
moment~\cite{ref:fox} of the event be less than 0.6.  We also require
that the cosine of the angle between the reconstructed $B$ direction
and the thrust axis of the rest of the event,
calculated in the \FourS rest frame, be less than 0.9.  These
criteria discriminate \FourS events from
non-resonant background events.

\Bz mesons are exclusively reconstructed by combining two charged \Dstar
candidates, using a number of \Dstarp and $D$ decay modes.
The \Dstarp mesons
are reconstructed in their decays \Dstptopip and \Dstptopiz. 
We include in this analysis the decay combinations \DstDst decaying 
to (\Dzpip, \Dzb\pim) or (\Dzpip, \Dmpiz), but not (\Dppiz,\Dmpiz) due to 
the smaller branching fraction and larger expected backgrounds.
\Dz and \Dp candidates are subjected to a mass-constrained fit to
provide an improved measurement of the $D$ meson's momentum. They are
combined with pion candidates, referred to as ``soft''
pions due to their low ($< 200\mevc$) transverse momentum, to form
\Dstarp candidates. A topological vertex fit is performed
that includes the mean position of the \epem interaction point
to improve the angular resolution of the soft pion.

The decay modes of the \Dz and \Dp 
are selected
by an optimization of \SsqovSpB based on Monte Carlo simulations,
where $S$ and $B$ are the expected number of signal and background
events, repsectively.
We first determine, based on Monte Carlo simulations, the expected
$S-B$ for each of the decay mode combinations individually.  Then, we
successively add modes in order of decreasing $S-B$ to compute an overall
\SsqovSpB value until \SsqovSpB no longer increases.
The decay modes used are $\Dz
\to \Km\pip$, $\Dz \to \Km\pip\piz$, $\Dz \to \Km\pip\pim\pip$, 
$\Dz \to \KS\pim\pip$,
$\Dp \to \Km\pip\pip$, $\Dp \to \KS\pip$, and $\Dp \to \Km\Kp\pip$.
\Dz (\Dp) meson candidates are required to
have a reconstructed invariant mass within 20\mevcc of the
nominal \Dz (\Dp) mass~\cite{pdg}.

Charged kaon candidates are required to be inconsistent with the pion
hypothesis, as inferred from the Cherenkov angle
measured by the DIRC and the ionizations measured by the SVT and DCH.  
No particle identification is required
for the kaon from the decay $\Dz \to \Km \pip$.

$\KS \to \pip\pim$ candidates are required to have an invariant mass
within 15\mevcc of the nominal \KS mass.  The angle between the
flight direction and the momentum vector of the \KS candidate is
required to be less than 200\mrad, and the transverse flight distance
from the primary event vertex, obtained from the remaining charged tracks in
the event, must be greater than 2\,mm.

Neutral pion candidates are formed from
pairs of photons in the EMC with energy above 30\mev,
an invariant mass within 35\mevcc\ of the nominal \piz\ mass,
and a summed energy greater than 200\mev.
A mass-constrained fit is applied to these \piz\ candidates. 
In the case of the soft \piz from \Dstptopiz decays,
the energy cut is replaced by a momentum cut, in the \FourS
frame, of $70 < p^* < 450\mevc$.

To select \Bz candidates with well reconstructed \Dstarp and $D$
mesons, we form a $\chi^2$ that includes all measured \Dstarp
and $D$ masses:

\begin{eqnarray}
\chisqM = 
   \left(\displaystyle \frac{m_D - \hat{m}_D}{\sigma_{m_D}}\right)^2
 + \left(\displaystyle
\frac{m_{\Db}-\hat{m}_{\Db}}{\sigma_{m_{\Db}}}\right)^2 \nonumber \\
   \mbox{}
 + \left(\displaystyle\frac{\Delta m_{D^{*}} - \Delta
   \hat{m}_{\Dstar}}{\sigma_{\Delta m_{D^{*}}}}\right)^2
 + \left(\displaystyle\frac{\Delta m_{\overline{D}^*} - \Delta
   \hat{m}_{\overline{D}^*}}{\sigma_{\Delta m_{\overline{D}^{*}}}}\right)^2,
\end{eqnarray}
where the caret over a value refers to the nominal value,
and $\Delta m_{D^{*}}$ 
is the $\Dstarp - D$ mass difference.  For $\sigma_{m_D}$ we use
values computed for each $D$ candidate, while for $\sigma_{\Delta m_{D^{*}}}$
we use fixed values of 0.83\mevcc for \Dstptopip and 1.18\mevcc
for \Dstptopiz.  A requirement that $\chisqM < 20$ is applied to all \Bz
candidates.  In events with more than one \Bz candidate, we choose the
candidate with the lowest value of \chisqM.

A $B$ meson candidate is characterized by two kinematic variables:
the energy-substituted mass,
\begin{equation}
\mes \equiv \sqrt{{E_{Beam}^{*2}} - {p_B^*}^2},
\end{equation}
and the difference of the $B$ candidate's energy from the beam energy,
\begin{equation}
\DeltaEStd \equiv E_{B}^* - E_{Beam}^*.
\end{equation}
$E_{B}^*$ ($p_B^*$) are the energy (momentum) of the \B\ candidate
in the center-of-mass frame and $E_{Beam}^*$ is one-half of the
total center-of-mass energy.
The signal region in the \DeltaEStd {\it vs.} \mes plane is defined to be
$|\DeltaEStd| < 25\mev$ and $5.273 < \mes < 5.285\gevcc$.
Based on Monte Carlo simulations,
the width of this region corresponds to approximately $\pm 2.5\sigma$
in both \DeltaEStd and \mes.

To determine the expected contribution from background in 
the signal region, we scale the number of events seen in a sideband in the
\DeltaEStd {\it vs.} \mes plane defined as
$|\DeltaEStd| < \DelEHiSide$, $\mesLowSide < \mes < \mesMidSide$
and
$\DelELowSide < |\DeltaEStd| < \DelEHiSide$, 
$5.26 < \mes < \mesHiSide$.
The scaling factor is calculated by parameterizing the shape of the 
background in the \DeltaEStd {\it vs.} \mes plane as the product of an ARGUS
function~\cite{ref:argus} in \mes and a first-order polynomial in \DeltaEStd.
Based on this parameterization we estimate that the ratio
of the number of background events in the signal region
to the number of events in the sideband region is
$(\fsideVal \pm \fsideErr)\times 10^{-2}$.
The uncertainty is derived from the observed variation of this ratio
under alternative assumptions 
for the background shape in \mes\ and \DeltaEStd using Monte Carlo simulations.
The simulations also indicate that there are no significant sources of
background appearing in the signal region beyond that indicated by
the sideband extrapolation.

\begin{figure}[!htb]
\begin{center}
\includegraphics[width=7.5cm]{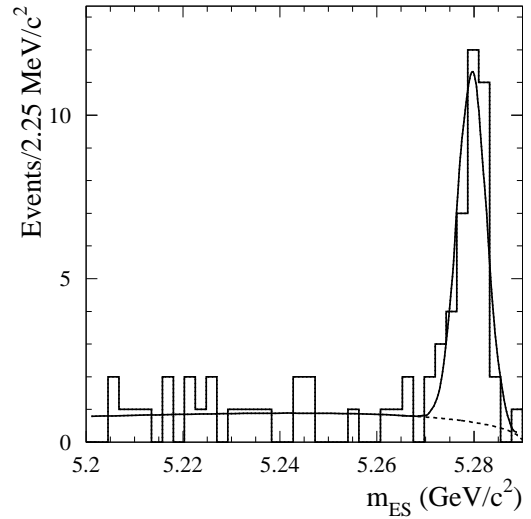}
\caption{The \mes istribution of \Bztodstdst\ events with
$|\DeltaEStd| < 25\,\mev$.  The curve represents a fit with
the sum of a Gaussian to model the signal and an ARGUS 
function~\cite{ref:argus} to model the background shape.
}
\label{fig:mes}
\end{center}
\end{figure}

After all selection criteria,
38 events are
located in the signal region, with 363 events in the sideband region.
The latter, together with the scaling factor determined above,
implies an expected number of
background events in the signal region of $6.24 \pm 0.33(stat) \pm
0.36 (syst)$.
The systematic uncertainty comes from the
background shape variation mentioned previously.  Figure~\ref{fig:mes}
shows a projection of the data on the \mes axis after requiring
$|\DeltaEStd| < 25\mev$.

We use a Monte Carlo simulation of the \babar\ detector to
determine the efficiency for reconstructing the signal.  The
efficiencies range from 17.4\% to 2.7\%, depending on the $D$ decay
modes. This, together with the total number of neutral $B$ mesons 
produced during data
collection, allows us to determine the branching fraction for
\Bztodstdst to be
$$\BRbztodstdst.$$

The high charged particle multiplicity makes this
measurement particularly sensitive to the
tracking system. Therefore
the dominant systematic uncertainty comes from our
level of understanding of the charged particle tracking efficiency.
Systematic errors 
are assigned on a per track basis for $\pi$, $K$, and soft $\pi$,
and are added linearly (9.9\%).  The effect on acceptance due to the 
imprecisely known partial-wave content of the \Bztodstdst\ final state
is another source of potential systematic bias (6.6\%). 
Other significant potential systematic biases arise due to the
uncertainties on the branching fractions~\cite{pdg} of the
${\Dstar}^+$, \Dz, and $D^+$ (5.6\%) and the uncertainties in mass
resolutions of reconstructed mesons (4.1\%). 
The total systematic uncertainty from all considered sources is 15\%. 


In addition to the branching fraction quoted above, we have also
measured the \CP-odd fraction of the final state.
This fraction, $R_\perp$, 
is determined from the angular distribution of the soft pions in
the decay, analyzed in the transversity basis~\cite{ref:dunietz}.  In
this reference frame, three decay amplitudes
determine the distribution of three decay angles.  Integrating
over time, $B$ flavor, and two of these three angles yields 
the following expression:
\begin{equation}
\frac{1}{\Gamma}\frac{\rm{d}\Gamma}{\rm{d}\cos\theta_{tr}} = 
\frac{3}{4}(1-R_\perp)\sin^2\theta_{tr} + 
\frac{3}{2}R_\perp\cos^2\theta_{tr}
.
\label{angdis}
\end{equation}
Here $\Gamma$ is the decay rate and 
$\theta_{tr}$ is the angle
between the normal to the ${\Dstar}^-$ decay plane and the line of flight
of the soft pion from the ${\Dstar}^+$ evaluated in the
${\Dstar}^+$ rest frame.  

We perform an unbinned maximum likelihood (ML) fit to the 38 events in the
signal region described previously.  The fit takes into
account the presence of background, whose properties are derived from
the sideband sample, and the angular resolution $\sigma_{\theta}$
estimated from Monte Carlo simulations.
We define the likelihood function to be
\begin{equation}
\begin{array}{r@{}c@{}l}
\displaystyle
{\cal{L}}=
\prod_{i=1,n}{\cal{L}}_i=
\prod_{i=1,n}
\left[p
\rule{0mm}{4mm}\right.
&\times&
\left.{\cal{F}}(\theta_{tr,i},\sigma_{\theta,i},R_\perp^{sig})+
\right.\\[5mm]
(1-p)
&\times&
\left.{\cal{F}}(\theta_{tr,i},\sigma_{\theta,i},R_\perp^{bkg})
\rule{0mm}{4mm}\right],
\end{array}
\label{ll1}
\end{equation}
where $n$ is the number of selected events. The contribution to the
total likelihood from the $i$-th event, ${\cal{L}}_i$, is defined 
in terms of the purity, $p$, of the sample and the 
probability density functions 
$\left.{\cal{F}}(\theta_{tr,i},\sigma_{\theta,i},R_\perp)\right.$
for the signal and background. 
$R_\perp^{sig}$ and $R_\perp^{bkg}$ are the parameters describing the shapes of the
signal and background angular distributions, respectively, and 
$\theta_{tr,i}$ is the measured transversity angle
in event $i$.
The probability density functions $\left.{\cal{F}}\right.$ 
are obtained from the
convolution of the angular distribution (Eq.~\ref{angdis}) with
Gaussian resolution functions describing the measurement uncertainties
$\sigma_{\theta,i}$.  From studies of simulated data,
$\sigma_{\theta}$ was measured to be 0.11 (0.12)\,radians for charged
(neutral) slow pions.  A 10\% uncertainty on these values is
considered when estimating the corresponding systematic error.

\begin{figure}[htb]
\begin{center}
\includegraphics[width=7.5cm]{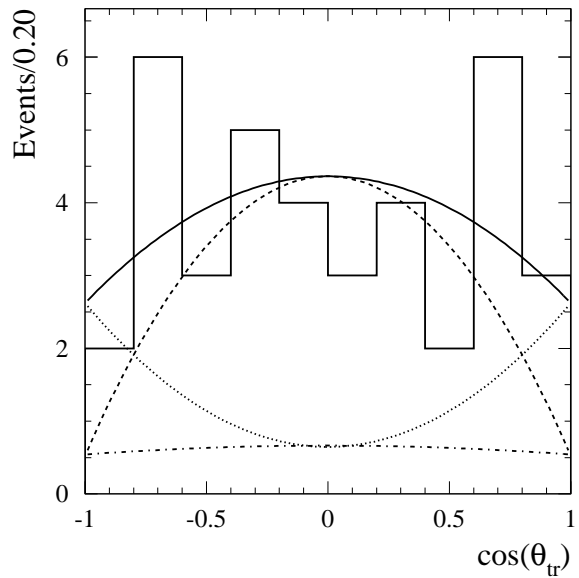}
\caption{The $\cos\theta_{tr}$ distribution from the unbinned ML fit,
superimposed on the histogram of the \Bztodstdst\ 
candidates in the signal region.
The solid  line represents the $\cos\theta_{tr}$ 
distribution from the unbinned ML fit for the selected events.
The dotted and dashed lines 
represent the fitted \CP-odd and \CP-even components, respectively,
for the signal.
The dot-dashed curve represents
the fitted background component. 
}
\label{fig:fit_sig}
\end{center}
\end{figure}

The value of $R_\perp^{bkg}$ is evaluated by fitting the 363 events in
the sideband region and setting $p=0$ in Eq.~\ref{ll1}.  The result
of this fit is
$R_\perp^{bkg}=0.29\pm0.04$, 
compatible with the value expected for a uniform distribution ($R_\perp=1/3$). 
To determine $R_\perp^{sig}$, we fit the 38 events in the signal region
with $R_\perp^{bkg}$ fixed to 0.29 and $p$ fixed at 83.6\%.  The result of
the fit to the signal region, without the correction for angular 
acceptance bias described below, is 
$R_\perp^{sig}=0.25\pm0.18(stat)$,
and is shown in Fig.~\ref{fig:fit_sig}.
The probability of obtaining a lower likelihood, evaluated using a
Monte Carlo technique, is 66\%.

It should be noted that Eq.~\ref{angdis} is the differential decay rate
$\Gamma$ integrated over the full ranges of the other two decay angles in 
the transversity basis, neglecting any bias in the projected
$\theta_{tr}$ distribution introduced by detector acceptance effects.
A detailed study of the kinematics of the decay shows that the
incomplete detector coverage of the polar angle with respect to the
beam axis does not introduce any bias in the distributions of the
decay angles in the transversity basis.  However, an
inefficiency in detecting soft pions below a threshold in transverse
momentum may indeed introduce such a bias due to the correlations
between the decay angles and particle momenta in the laboratory
frame.  The amount of these acceptance losses depends on the
population of phase space, determined by the values of the decay
amplitudes.

An accurate correction for these acceptance effects requires the
complete determination of the decay amplitudes using a full angular
analysis on a sufficiently large data sample.  To estimate the size of
the acceptance bias on $R_\perp$ without knowing the decay amplitudes,
the fit procedure was tested on several samples of \Bztodstdst
simulated events generated with different sets of decay amplitudes.
The different amplitudes affect to varying extents the correlated soft pions'
transverse momenta and angular distributions.
The fitted $R_\perp$ values were found to be consistent with the
generated values
in the limit of negligible soft pion inefficiency. 
Depending on the mix of decay amplitudes, they did reveal a bias once
the pion-detection threshold 
was taken into account. Considering the full possible range of decay
amplitudes, the calculated bias on $R_\perp$ ranged from $-0.048$ to
$+0.004$.
The central value of this interval is taken 
as a correction to the fitted $R_\perp^{sig}$, while its half width is
taken as 
an estimate of the corresponding systematic uncertainty (0.026).
Additional, smaller systematic uncertainties affecting the $R_\perp$ 
measurement arise from the imperfect knowledge of 
the resolution in the transversity angle $\theta_{tr}$ (0.006), 
the angular distribution of the background (0.008), 
and the purity of the signal sample (0.0003).  
The total systematic uncertainty on
$R_\perp$ is determined to be 0.028, giving the final corrected
result:
$$R_\perp = 0.22 \pm 0.18(stat) \pm 0.03(syst).$$


In summary,
we have observed a signal of $31.8 \pm 6.2(stat) \pm 0.4(syst)$ events in
the decay \Bztodstdst.  Our measurement of the branching ratio is
$$\BRbztodstdst.$$
From the transversity angular distribution of these events,
we have also measured the \CP-odd fraction, $R_\perp$,
of the final state.
These measurements provide a starting point for
measuring time-dependent \CP-violating asymmetries in these
decays when more data become available.

We are grateful for the excellent luminosity and machine conditions
provided by our \pep2\ colleagues, 
and for the substantial dedicated effort from
the computing organizations that support \babar.
The collaborating institutions wish to thank 
SLAC for its support and kind hospitality. 
This work is supported by
DOE
and NSF (USA),
NSERC (Canada),
IHEP (China),
CEA and
CNRS-IN2P3
(France),
BMBF
(Germany),
INFN (Italy),
NFR (Norway),
MIST (Russia), and
PPARC (United Kingdom). 
Individuals have received support from the 
A.~P.~Sloan Foundation, 
Research Corporation,
and Alexander von Humboldt Foundation.


\begin{thebibliography}{99}
\bibitem{ref:cc}
Charge-conjugate states are implied throughout this paper and the symbol
$D^{(*)}$ refers to either $D$ or $D^*$.
\end{thebibliography}
\end{document}